\documentclass[12pt]{iopart}

\usepackage{amsmath}
\usepackage{graphicx}
\usepackage{cite}
\usepackage[usenames,dvipsnames,svgnames,table]{xcolor}
\usepackage{hyperref}
\hypersetup{
    colorlinks,
    linkcolor={red!50!black},
    citecolor={blue!50!black},
    urlcolor={blue!80!black}
}

\begin{document}

\title{Plasmonic excitations in double-walled carbon nanotubes}

\author{Pablo Mart\'in-Luna$^1$, Alexandre Bonatto$^2$, Cristian Bontoiu$^3$, Guoxing Xia$^4$, Javier Resta-L\'opez$^5$}

\address{$^1$ Instituto de F\'isica Corpuscular (IFIC), Universitat de Val\`encia - Consejo Superior de Investigaciones Cient\'ificas, 46980 Paterna, Spain}

\address{$^2$ Graduate Program in Information Technology and Healthcare Management, and the Beam Physics Group, Federal University of Health Sciences of Porto Alegre, Porto Alegre, RS, 90050-170, Brazil}

\address{$^3$ Department of Physics, The University of Liverpool, Liverpool L69 3BX, United Kingdom, \\ The Cockcroft Institute, Sci-Tech Daresbury, Warrington WA4 4AD, United Kingdom}

\address{$^4$ Department of Physics and Astronomy, The University of Manchester, Manchester M13 9PL, United Kingdom \\ 
The Cockcroft Institute, Sci-Tech Daresbury, Warrington WA4 4AD, United Kingdom}

\address{$^5$ Instituto de Ciencia de los Materiales (ICMUV), Universidad de Valencia, 46071 Valencia, Spain}

\ead{pablo.martin@uv.es, javier2.resta@uv.es}

\vspace{10pt}
\begin{indented}
\item[]Keywords: double-walled carbon nanotubes, wakefield, hydrodynamic model, plasmons \\

\item[]\today
\end{indented}

\begin{abstract}
The interactions of charged particles moving paraxially in multi-walled carbon nanotubes (MWCNTs) may excite electromagnetic modes. This wake effect has recently been proposed as a potential novel method of short-wavelength high-gradient particle acceleration. In this work, the excitation of  wakefields in double-walled carbon nanotubes (DWCNTs) is studied by means of the linearized hydrodynamic theory. General expressions have been derived for the excited longitudinal and transverse wakefields and related to the resonant wavenumbers which can be obtained from the dispersion relation. In the absence of friction, the stopping power of the wakefield driver, modelled here as a charged macroparticle, can be written solely as a function of these resonant wavenumbers. The dependencies of the wakefields on the radii of the DWCNT and the driving velocity have been studied. DWCNTs with inter-wall distances much smaller than the internal radius may be a potential option to obtain higher wakefields for particle acceleration compared to single-walled carbon nanotubes (SWCNTs).


\end{abstract}

\maketitle


\section{Introduction}\label{Introduction}

Since the discovery of carbon nanotubes (CNTs) by S. Iijima in 1991 \cite{iijima1991CNTdiscovery}, they have attracted great interest for applications in different areas of research and technology. CNTs can exhibit metallic or semiconductor properties depending on their rolling pattern and have unique thermo-mechanical and electronic properties as well as dimensional flexibility. For this reason, CNTs have been widely studied in both theoretical and experimental aspects. For instance, CNTs could be used for channeling and steering charged particles similar to crystal channeling due to their hollow structure. In particular, experimental results on 300 keV electrons \cite{Chai2007_channel300kev_electrons} and 2 MeV $\text{He}^{+}$ ions \cite{Zhu2005_SPIE_channel_ion} channeling in CNTs have been reported.


On the other hand, in the 1980s and 1990s T. Tajima and others \cite{TajimaCavenago1987_XrayAccelerator_PhysRevLett.59.1440, chen1987solid, chen1997crystal} proposed the solid-state wakefield acceleration using crystals as a potential particle acceleration technique to sustain TV/m acceleration gradients. In the original Tajima’s conceptual scheme \cite{TajimaCavenago1987_XrayAccelerator_PhysRevLett.59.1440}, high energy ($\approx$40 keV) X-rays are injected into a crystalline lattice at the Bragg angle to cause Borrmann-Campbell effect \cite{borrmann1950absorption, campbell1951x}, generating a longitudinal electric wakefield which can be used as an accelerating structure.  Similarly, ultrashort charged particle bunches can excite electric wakefields so that the energy loss of the driving bunch can be transformed into an increment of energy for
a properly injected witness bunch. However, the beam intensity acceptance and the dechanneling rate are limited due to the angstrom-size channels of natural crystals. In this context, CNTs might be an alternative candidate for TeV/m acceleration as they have wider channels in two dimensions, longer dechanneling lengths \cite{BIRYUKOV2002_PLB_channelCNTs, BELLUCCI2005_PLB_channelCNT}, larger degree of dimensional flexibility and thermo-mechanical strength. For this reason, carbon-based nanostructures such as CNTs and graphene layers are currently being investigated for wakefield acceleration \cite{Bonatto2023_effective_plasma_POP, Bontoiu2023_catapult, Martin-Luna2023_ExcitationWakefieldsSWCNT_NJP}.

Wakefields in CNTs can be excited through the collective oscillation of the free electron gas confined over the nanotubes surfaces (often referred to as plasmons), which is triggered by the driving bunch. Electronic excitations on SWCNTs or MWCNTs surfaces have been theoretically studied using a dielectric theory \cite{Arista2001IonsPhysRevA.64.032901, Arista2001ChargedParticlesPhysRevB.63.165401, WangMiskovic2002_dielectric_theory_PhysRevA.66.042904}, a hydrodynamic model \cite{Stockli2001PhysRevB.64.115424, WangMiskovic2004_Hydro_theory_PhysRevA.69.022901, MOWBRAY2004_hydrodynamic_model_wake_effects, MOWBRAY2005_MWCNT}, a two-fluid model \cite{WangMiskovic2004TwoFluidModelPhysRevB.70.195418, CHUNG2007_MWCNT_theory_RadPhysChem}, a quantum hydrodynamic model \cite{WeiWang2007_quantum_hydrodynamic_model_PhysRevB.75.193407}, a kinetic model \cite{Song2006_KineticModel_PhysRevA.78.012901, Zhao2008_ChinPhysLett_KineticModel, YOU2009_KineticModelDWCNT_NIMB, ZHANG2014_kinetic_model_Carbon} and even a combination of a semi-classical kinetic model with the molecular dynamics method \cite{Zhang2017_TWCNT_EurPhysJD}. These previous studies mostly explore properties such as the energy loss and stopping power or, at most, evaluate the induced surface electron density and/or induced potential \cite{MOWBRAY2004_hydrodynamic_model_wake_effects, Zhang2013_WakeEffects_ChinPhysLett_KineticModel, ZHANG2014_kinetic_model_Carbon}. However, they do not address the induced longitudinal and transverse wakefields, which could provide acceleration and focusing, respectively, for a witness charge.


In a recent work \cite{Martin-Luna2023_ExcitationWakefieldsSWCNT_NJP} we have studied the wakefields excited by charged particles moving paraxially inside a SWCNT using the hydrodynamic model, providing analytical expressions that enable the rapid optimization of the SWCNT parameters to achieve the maximum longitudinal wakefield.
The hydrodynamic model was chosen because of its simplicity and good agreement with the dielectric formalism in random-phase approximation \cite{WangMiskovic2004_Hydro_theory_PhysRevA.69.022901}. Furthermore, this model
explains the modifications inflicted on SWCNTs irradiated with swift heavy ions \cite{MOSLEM2022_SWCNTmodifications_ResultsInPhysics105438}.

As layered materials exhibit very
rich spectra of plasmonic excitations \cite{Yannouleas1996_DispersionLayeredCarbonStructures_PhysRevB.53.10225}, the motivation of the present work is to extend our previous work to DWCNTs, comparing the results with SWCNTs. In this context, targets with multiple concentric CNT arrays, which could be interpreted as MWCNTs, are being investigated \cite{Bonatto2023_effective_plasma_POP}. In this article, we have related the excited wakefields in DWCNTs with the resonant wavenumbers obtained from the dispersion relation, which splits in two branches compared to the case of a SWCNT. It is worth noting that the expressions that we have obtained can be used to calculate the stopping power only in terms of the resonant wavenumbers if the frictional force on the delocalized electrons of the carbon ions due to scattering on the positive-charge background is negligible.

This article is organized as follows. In Section \ref{Linearized hydrodynamic theory} the linearized hydrodynamic theory is reviewed for a MWCNT and general expressions are derived for the longitudinal and transverse wakefields excited by the interaction of a charged particle with the MWCNT. These expressions are further particularized for the case of a DWCNT. For simplicity, atomic units are used during the derivation of the wakefield expressions, which are later converted to SI units to present the final results. In Section \ref{Results and discussion}, the influence of different model parameters is investigated, relating them with the dispersion relation. Finally, the main conclusions of this work are presented in Section \ref{Conclusions}.

\section{Linearized hydrodynamic theory}\label{Linearized hydrodynamic theory}

\begin{figure}[!h]
\includegraphics[width=\columnwidth]{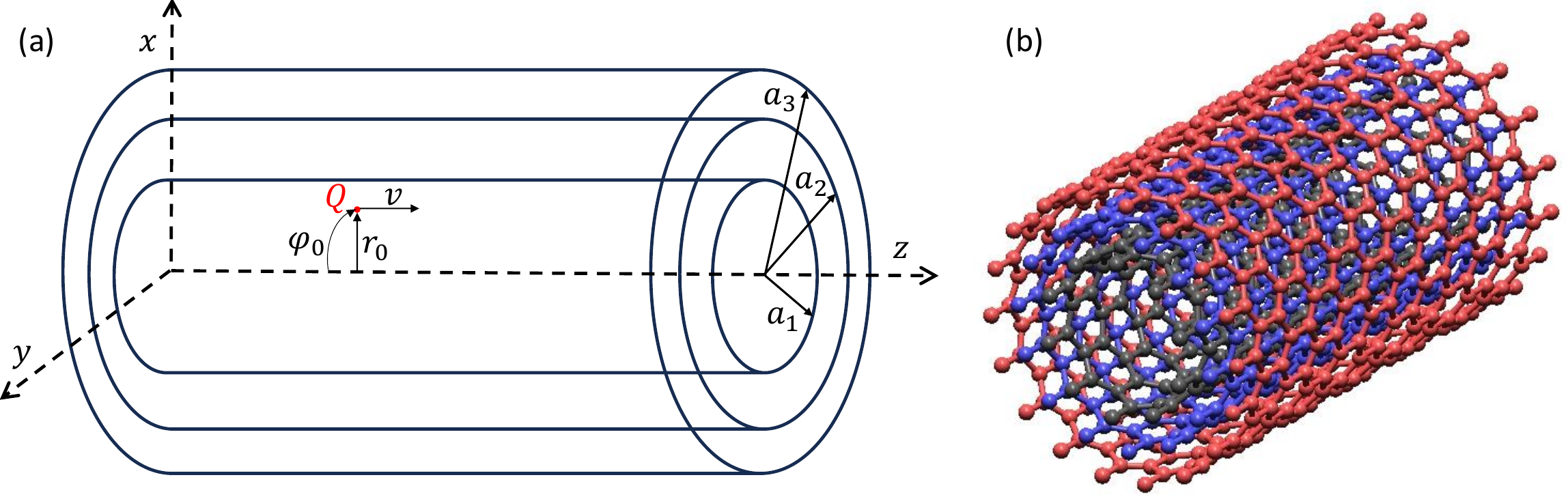}
\centering
\caption{(a) Scheme of the considered charge $Q$ travelling parallel to the $z$-axis inside a MWCNT with $N=3$ cylinders. (b) Schematic model of the hexagonal lattice of the cylinder walls with black, blue and red represent the inner, middle and outer layer respectively.}
\label{fig:CNT_scheme}
\end{figure}

In this work, a linearized hydrodynamic theory is adopted to describe excitations produced by a single charged particle on the nanotube surfaces of a DWCNT. We start by reviewing the general case of a MWCNT composed of $N$ concentric cylinders with radii $a_1<a_2<\dotsi<a_N$ \cite{MOWBRAY2005_MWCNT} which will be analyzed in detail for the case $N=2$. In this model, each wall is modelled as an infinitesimally thin and infinitely long cylindrical shell. The delocalized electrons of the carbon ions are considered as a two-dimensional free-electron gas (Fermi gas) that is confined over the cylindrical surfaces of the MWCNT with a uniform surface density $n_0$. A driving point-like charge $Q$, travelling parallel to the $z$-axis inside the MWCNT with a constant velocity $v$, is considered (see Fig. \ref{fig:CNT_scheme}). Hence, its position as a function of time $t$ is $\mathbf{r}_0(t)=\left(r_0, \varphi_0, vt\right)$ in cylindrical coordinates. The homogeneous electron gas at each wall, which will be perturbed by the driving charge Q, can be modelled as a charged fluid, with velocity fields $\mathbf{u}_j(\mathbf{r}_j ,t)$ and surface density $n(\mathbf{r}_j ,t)=n_0+n_j(\mathbf{r}_j ,t)$, where $\mathbf{r}_j=(a_j, \varphi,z)$ are the coordinates of a point at the cylindrical surface of the $j$th tube, and $n_j(\mathbf{r}_j ,t)$ is its perturbed density per unit area. In the linearized hydrodynamic model, it is assumed that the perturbed densities $n_j$ and the fluid velocities $\mathbf{u}_j$ are relatively small perturbations. As the electron gases are confined to the cylindrical surfaces, the normal component to the surface of the tubes of the velocity fields $\mathbf{u}_j$ is zero. The time scale associated with the ionic motion is orders of magnitude slower than that of the electronic motion since the carbon ions are much heavier than the electrons. Therefore, for the purpose of investigating the wakefield dynamics, the ionic motion can be neglected \cite{Hakimi2018_ionic_motion_PoP_10.1063/1.5016445, Hakimi2020_ionic_motion_nanotube_doi:10.1142/S0217751X19430115}.

In the linearized hydrodynamic model, the electronic excitations on the tube wall can be described by the continuity equation

\begin{equation}\label{eq:continuity_equation}
\frac{\partial n_j\left(\mathbf{r}_j, t\right)}{\partial t}+n_0 \nabla_{j} \cdot \mathbf{u}_j\left(\mathbf{r}_j, t\right)=0,
\end{equation}



\noindent and the momentum-balance equation of the electron fluid at each tube surface

\begin{equation}\label{eq:momentum_balance_equation}
\begin{split}
\frac{\partial \mathbf{u}_j\left(\mathbf{r}_j, t\right)}{\partial t}=\nabla_{j} \Phi\left(\mathbf{r}_j, t\right)-\frac{\alpha}{n_0} \nabla_{j} n_j\left(\mathbf{r}_j, t\right)+\frac{\beta}{n_0} \nabla_{j}\left[\nabla_{j}^2 n_j\left(\mathbf{r}_j,t\right)\right]-\gamma \mathbf{u}_j\left(\mathbf{r}_j, t\right),
\end{split}
\end{equation}


\noindent where we have retained only the first-order terms in $n_j$ and $\mathbf{u}_j$. In these equations, $\mathbf{r}=(r,\varphi,z)$ is the position vector, $\nabla=\hat{\mathbf{r}}\frac{\partial}{\partial r} +\hat{\boldsymbol{\varphi}}\frac{1}{r} \frac{\partial}{\partial \varphi} +\hat{\mathbf{z}}\frac{\partial}{\partial z} $,  $\nabla_j=\hat{\boldsymbol{\varphi}}\frac{1}{a_j} \frac{\partial}{\partial \varphi} +\hat{\mathbf{z}}\frac{\partial}{\partial z}$ differentiates only tangentially to the $j$th tube surface and $\Phi$ is the electric scalar potential. Equation (\ref{eq:momentum_balance_equation}) shows the sum of four different contributions. The first term in the right-hand side is the force on electrons which belong to the $j$th nanotube surface due to the tangential component of the electric field generated by the driving charge $Q$ and all the consequent perturbed densities. The second term considers the possible coupling with acoustic modes defining the parameter $\alpha=v^2_F/2$ (in which $v_F=(2\pi n_0)^{1/2}$ is the Fermi velocity of the two-dimensional electron gas), and the third term is a quantum correction that arises from the functional derivative of the Von Weizsacker gradient correction in the equilibrium kinetic energy of the electron fluid \cite{Nejati2009_doi:10.1063/1.3077306} and describes single-electron excitations in the electron gas, where it has been defined the parameter $\beta=\frac{1}{4}$. The last term represents a frictional force on electrons due to scattering with the ionic-lattice charges, where $\gamma$ is the damping parameter. This friction parameter may be also used as a phenomenological parameter to take into account the broadening of the plasmon resonance in the excitation spectra of different materials \cite{Arista2001IonsPhysRevA.64.032901}.
The equations (\ref{eq:continuity_equation})-(\ref{eq:momentum_balance_equation}) are coupled by the 3D Poisson's equation in free space. Therefore, the total electric potential is $\Phi=\Phi_{\mathrm{0}}+\Phi_{\mathrm{ind}}$, where $\Phi_0=\frac{Q}{\left\|\mathbf{r}-\mathbf{r}_{\mathbf{0}}\right\|}$ is the Coulomb potential generated by the driving charge and $\Phi_{\mathrm{ind}}$ is the potential created by the perturbation of the electron fluids:

\begin{equation}\label{eq:Phi_ind}
\Phi_{\text {ind }}(\mathbf{r}, t)=-\sum_j \int \mathrm{d}^2 \mathbf{r}_j^{\prime} \frac{n_j\left(\mathbf{r}_j^{\prime}, t\right)}{\left\|\mathbf{r}-\mathbf{r}_j^{\prime}\right\|},
\end{equation}

\noindent where $\mathbf{r}_j^{\prime}=(a_j,\varphi^{\prime},z^{\prime})$ are the cylindrical coordinates of a generic point at the surface of the $j$th tube and $ \mathrm{d}^2 \mathbf{r}_j^{\prime}=a_j\mathrm{d}\varphi^{\prime}\mathrm{d}z^{\prime}$.

In order to solve the equations (\ref{eq:continuity_equation})-(\ref{eq:Phi_ind}) we define the Fourier-Bessel (FB) transform $\widetilde{A}(m, k, \omega)$ of an arbitrary function $A(\varphi, z, t)$ by

\begin{equation}\label{eq:FB}
A(\varphi, z, t)=  \sum_{m=-\infty}^{\infty} \int_{-\infty}^{\infty} \frac{\mathrm{d} k}{(2 \pi)^2}
 \int_{-\infty}^{\infty} \frac{\mathrm{d} \omega}{2 \pi} \mathrm{e}^{i k z+i m \varphi-i \omega t} \widetilde{A}(m, k, \omega).
\end{equation}

\noindent where the index $m$ denotes the different angular-momentum modes. In particular, the Coulomb potential can be expressed as

\begin{equation}\label{eq:FB_Coulomb}
\frac{1}{\left\|\mathbf{r}-\mathbf{r}^{\prime}\right\|}=\sum_{m=-\infty}^{\infty} \int_{-\infty}^{\infty} \frac{\mathrm{d} k}{(2 \pi)^2} \mathrm{e}^{i k\left(z-z^{\prime}\right)+i m\left(\varphi-\varphi^{\prime}\right)} g\left(r, r^{\prime} ; m, k\right),
\end{equation}

\noindent where $g\left(r, r^{\prime} ; m, k\right) \equiv 4 \pi I_m\left(|k| r_{\text{min}}\right) K_m\left(|k| r_{\text{max}}\right)$ with $r_{\text{min}}=\text{min}(r,r^{\prime})$, $r_{\text{max}}=\text{max}(r,r^{\prime})$ and $I_m(x)$ and $K_m(x)$ are the modified Bessel functions of integer order $m$. Substituting the relation (\ref{eq:FB_Coulomb}) in (\ref{eq:Phi_ind}), the FB transform of $\Phi_{\text {ind }}$ is obtained as

\begin{equation}\label{eq:Phi_ind_FB}
\tilde{\Phi}_{\text {ind }}(r, m, k, \omega)=-\sum_j g\left(r, a_j ; m, k\right) a_j \tilde{n}_j(m, k, \omega).
\end{equation}

\noindent where $\tilde{n}_j(m, k, \omega)$ is the FB of the perturbed density $n_j$. After eliminating $\mathbf{u}_j$ in (\ref{eq:momentum_balance_equation}) by using the continuity equation and applying the FB definition, the FB transform of the perturbed densities are related by the system of $N$ coupled linear equations

\begin{equation}\label{eq:matrix_n}
S_j(m,k,\omega)\tilde{n}_j(m, k, \omega)-\sum_l G_{j l}(m, k)\tilde{n}_l(m, k, \omega)=B_j(m,k,\omega),
\end{equation}

\noindent where we have defined the following functions

\begin{equation}\label{eq:S_j}
S_j(m,k,\omega)=\omega(\omega+i\gamma)-\alpha\left(k^2+\frac{m^2}{a_j^2}\right)-\beta\left(k^2+\frac{m^2}{a_j^2}\right)^2,
\end{equation}

\begin{equation}\label{eq:G_jl}
G_{j l}(m, k) = n_0 a_l\left(k^2+\frac{m^2}{a_j^2}\right) g\left(a_j, a_l ; m, k\right),
\end{equation}

\begin{equation}\label{B_j}
B_j(m,k,\omega)=-n_0\left(k^2+\frac{m^2}{a_j^2}\right) \tilde{\Phi}_0\left(a_j, m, k, \omega\right).
\end{equation}

\noindent The FB transform of the Coulomb potential created by the driving charge is

\begin{equation}\label{eq_Phi_0}
\tilde{\Phi}_{0}(r, m, k, \omega)=2 \pi Q g\left(r, r_0 ; m, k\right) \delta(\omega-k v) \exp \left(-i m \varphi_0\right).
\end{equation}

Thus, the induced longitudinal and transverse electric wakefields are, respectively, 

\begin{equation}\label{eq:Wz}
\begin{split}
W_{z,\mathrm{ind}}(r, \varphi, \zeta) =-\frac{\partial \Phi_{\text{ind}}}{\partial z} &=\frac{1}{(2 \pi)^3} \sum_{m=-\infty}^{+\infty}e^{i m\varphi} \int_{-\infty}^{+\infty} \mathrm{d} k\, k \left(\text{Re}\left[\tilde{\Phi}_{\text {ind}}(r, m, k, kv)\right] \sin (k \zeta)\right. \\ 
&+ \text{Im}\left[\tilde{\Phi}_{\text {ind}}(r, m, k, kv)\right] \cos (k \zeta) )=W_{z,\text{Re}}+W_{z,\text{Im}},
\end{split}
\end{equation}

\begin{equation}\label{eq:Wr}
\begin{split}
W_{r,\mathrm{ind}}(r, \varphi, \zeta) =-\frac{\partial \Phi_{\text{ind}}}{\partial r} &=-\frac{1}{(2 \pi)^3} \sum_{m=-\infty}^{+\infty}e^{i m\varphi} \int_{-\infty}^{+\infty} \mathrm{d} k\, \left(\text{Re}\left[\partial_r \tilde{\Phi}_{\text {ind}}(r, m, k, kv)\right] \cos (k \zeta)\right. \\ 
&- \text{Im}\left[\partial_r\tilde{\Phi}_{\text {ind}}(r, m, k, kv)\right] \sin (k \zeta) )=W_{r,\text{Re}}+W_{r,\text{Im}},
\end{split}
\end{equation}

\noindent where a comoving coordinate $\zeta=z-vt$ has been defined and the following properties:
$\text{Re}\left[\tilde{f}(k, kv)\right]=\text{Re}\left[\tilde{f}(-k, -kv)\right]$, $\text{Im}\left[\tilde{f}(k, kv)\right]=-\text{Im}\left[\tilde{f}(-k, -kv)\right]$ were applied for $\tilde{f}(k, kv)=\{\tilde{\Phi}_{\text {ind}}(r, m, k, kv),\partial_r\tilde{\Phi}_{\text {ind}}(r, m, k, kv)\}$; $\text{Re}$ and $\text{Im}$ denote the real and imaginary part, respectively.
The previous integrals have been separated in two terms, which come from $\text{Re}[\tilde{\Phi}_{\text{ind}}]$ and $\text{Im}[\tilde{\Phi}_{\text{ind}}]$, respectively. To reduce the computational time and prevent artificial numerical errors, a cutoff for large wavenumbers $k$ has to be introduced in the numerical integration. 
It is worth noting that Eq. (\ref{eq:matrix_n}) can be expressed as the following matrix equation

\begin{equation}\label{eq_M_matrix}
M\vec{\tilde{n}}=\vec{B}, \quad \quad M_{jj}=S_j-G_{jj}, \quad M_{ij}=-G_{ij} \quad (i \neq j).
\end{equation}

Consequently, the resonant frequencies of the collective excitations in the coupled 2D fluids can be calculated by solving the eigenvalue equation $\det(M)=0$ for $\gamma=0$. In particular, if $N=2$ the eigenvalue equation gives the following dispersion relations: 

\begin{equation}\label{eq:dispersion_w_pm}
\omega_{ \pm}^2(m,k)=\frac{\omega_1^2+\omega_2^2}{2} \pm \sqrt{\left(\frac{\omega_1^2-\omega_2^2}{2}\right)^2+\Delta^2},
\end{equation}

\noindent where

\begin{equation}\label{eq:dispersion_wj}
\omega_{j}^2=\alpha\left(k^2+m^2 / a_j^2\right)+\beta\left(k^2+m^2 / a_j^2\right)^2+n_0 a_j\left(k^2+m^2 / a_j^2\right) g\left(a_j, a_j ; m, k\right)
\end{equation}

\noindent are the dispersion relations of the individual electron
fluids on the tubes $j=1,2$ and 

\begin{equation}\label{eq:Delta}
\Delta^2=n_0^2 a_1 a_2\left(k^2+m^2 / a_1^2\right)\left(k^2+m^2 / a_2^2\right) g^2\left(a_1, a_2 ; m, k\right)
\end{equation}

\noindent describes the electrostatic interaction between both fluids. Thus, in the case of DWCNTs, the solution of (\ref{eq:matrix_n}) is

\begin{equation}\label{eq:n_FB}
\tilde{n}_j(m, k, \omega)=\frac{N_j(m,k,\omega)}{D_m^{+}(k,\omega)D_m^{-}(k,\omega)}\delta(\omega-k v),
\end{equation}

\noindent where 
\begin{equation}\label{eq:N}
N_1(m,k,\omega)=(S_2-G_{22})B^{\prime}_1+G_{12}B^{\prime}_2, \quad N_2(m,k,\omega)=(S_1-G_{11})B^{\prime}_2+G_{21}B^{\prime}_1, 
\end{equation}

\begin{equation}\label{eq:D_pm}
D_m^{\pm}(k,\omega)=\omega(\omega+i \gamma)-\omega_{\pm}^2(m,k),
\end{equation}

\noindent and $B^{\prime}_j$ are the functions $B_j$ without the Dirac delta $\delta(\omega-k v)$, i.e.

\begin{equation}\label{eq:B_prime}
B^{\prime}_j(m,k)\delta(\omega-k v)=B_j(m,k,\omega).
\end{equation}

It is important to note that the terms $W_{z,\text{Im}}$ and $W_{r,\text{Im}}$ can be analytically integrated if the damping factor vanishes ($\gamma \rightarrow 0^+$): 

\begin{equation}\label{eq:Wz_im}
W_{z,\text{Im}}(r, \varphi, \zeta) =\sum_{m=-\infty}^{+\infty}e^{i m\varphi}[W_{z,m}^+\cos(k_m^+\zeta)+W_{z,m}^-\cos(k_m^-\zeta)],
\end{equation}

\begin{equation}\label{eq:Wr_im}
W_{r,\text{Im}}(r, \varphi, \zeta) = \sum_{m=-\infty}^{+\infty}e^{i m\varphi}[W_{r,m}^+\sin(k_m^+\zeta)+W_{r,m}^-\sin(k_m^-\zeta)],
\end{equation}

\begin{equation}\label{eq:Wz_mas}
W_{z,m}^+=\frac{k_m^+}{(2 \pi)^2} \left(D_m^{-}(k_m^+,k_m^+v)\left|\frac{\partial Z_m^+}{\partial k}\right|_{k=k_m^+}\right)^{-1} \sum_{j=1}^2 g\left(r, a_j ; m, k_m^+\right)N_j(m,k_m^+,k_m^+v),
\end{equation}

\begin{equation}\label{eq:Wz_menos}
W_{z,m}^-=\frac{k_m^-}{(2 \pi)^2} \left(D_m^{+}(k_m^-,k_m^-v)\left|\frac{\partial Z_m^-}{\partial k}\right|_{k=k_m^-}\right)^{-1} \sum_{j=1}^2 g\left(r, a_j ; m, k_m^-\right)N_j(m,k_m^-,k_m^-v),
\end{equation}

\begin{equation}\label{eq:Wr_mas}
W_{r,m}^+=\frac{1}{(2 \pi)^2}\left(D_m^{-}(k_m^+,k_m^+v)\left|\frac{\partial Z_m^+}{\partial k}\right|_{k=k_m^+}\right)^{-1} \sum_{j=1}^2 \partial_r g\left(r, a_j ; m, k_m^+\right)N_j(m,k_m^+,k_m^+v),
\end{equation}

\begin{equation}\label{eq:Wr_menos}
W_{r,m}^-=\frac{1}{(2 \pi)^2}\left(D_m^{+}(k_m^-,k_m^-v)\left|\frac{\partial Z_m^-}{\partial k}\right|_{k=k_m^-}\right)^{-1} \sum_{j=1}^2 \partial_r g\left(r, a_j ; m, k_m^-\right)N_j(m,k_m^-,k_m^-v),
\end{equation}

\noindent where the quantity $Z_m^{\pm}(k)=\text{Re}[D_m^{\pm}(k,kv)]=(k v)^2-\omega_{\pm}^2(m,k)$ has been defined, and $k_m^{\pm}$ are the positive roots of $Z_m^{\pm}(k)$, i.e. the condition of the plasma resonance $k_m^{\pm}v=\omega_{\pm}(m,k)$. For completeness, it is worth noting that the stopping power $\left.\left.S = -Q W_z \right|_{\mathbf{r}=\mathbf{r}_0}=-Q W_{z,\text{Im}}\right|_{\mathbf{r}=\mathbf{r}_0}$ (that is the energy loss of a channelled particle per unit path length due to the collective electron excitations on the nanotube walls) can be calculated using the expression (\ref{eq:Wz_im}) when the damping factor vanishes.

\section{Results and discussion} \label{Results and discussion}

As it can be deduced from Equations (\ref{eq:Wz_im})-(\ref{eq:Wr_menos}), the roots $k_m^{\pm}$ given by the plasma resonance are essential to describe the behaviour of the wakefields. For this reason, this section begins with a detailed analysis of the dispersion relation. In the following calculations, unless otherwise indicated, it is assumed that the surface electron density of each wall can be approximated by the electron-gas density of a graphite sheet: $n_0=n_g=1.53\times10^{20}$\,m$^{-2}$ \cite{Ostling1997_surface_density_PhysRevB.55.13980, WangMiskovic2004_Hydro_theory_PhysRevA.69.022901}.

\subsection{Dispersion relation}\label{Dispersion relation}

\begin{figure}[!h]
\includegraphics[width=\columnwidth]{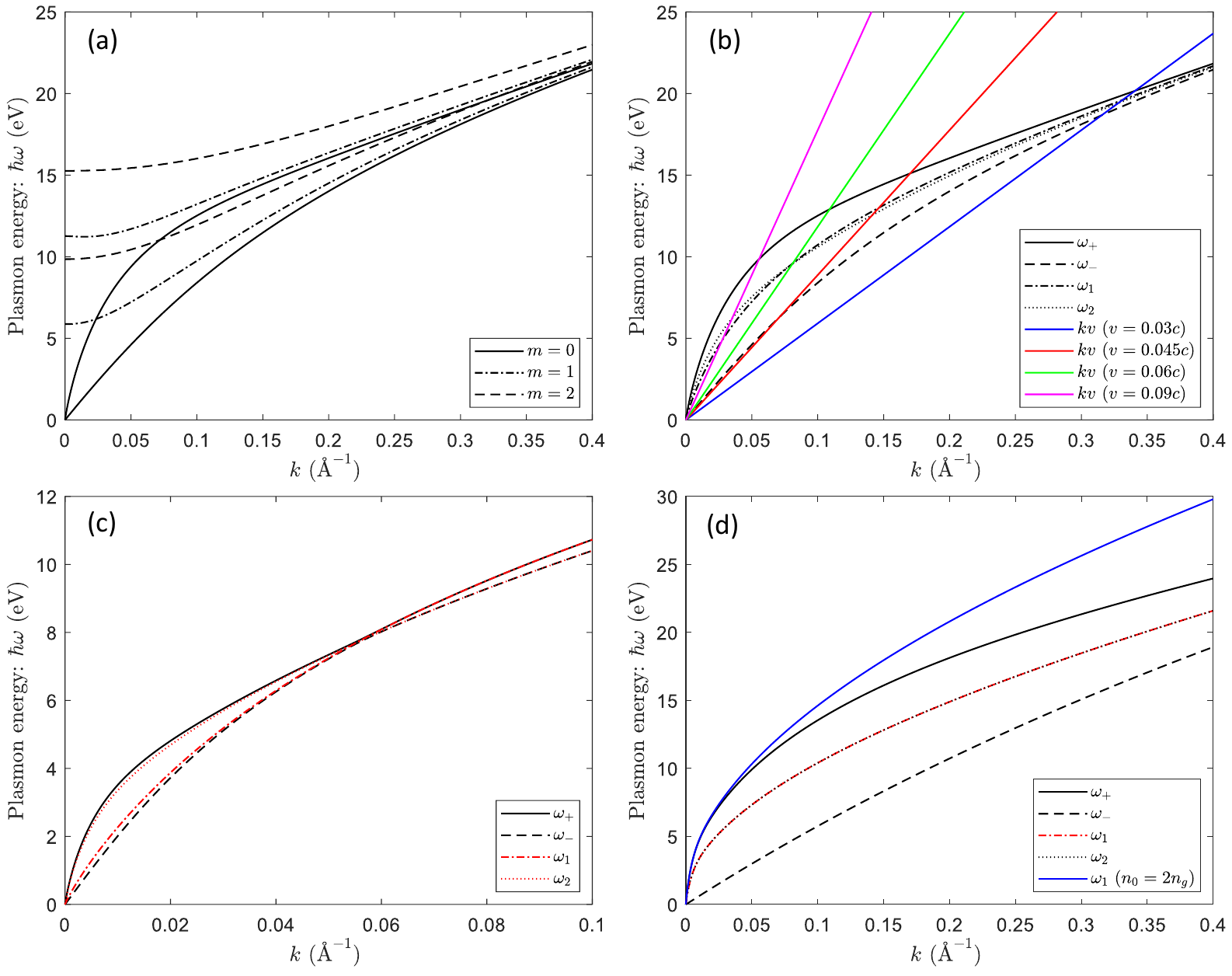}
\centering
\caption{(a) Dispersion curves $\omega_{\pm}(m,k)$ for several angular-momentum modes for $a_1=1$\,nm and $a_2=2$\,nm. (b) Fundamental modes $m=0$ for $a_1=1$\,nm and $a_2=2$\,nm compared to the individual electron fluid frequencies $\omega_1$ and $\omega_2$. The resonances $k_0^{\pm}$ are the intersection of the $kv$ lines (plotted for $v=0.03c, v=0.045c, v=0.06c$ and $v=0.09c$) with the dispersion curves. Fundamental modes $m=0$ for (c) $a_1=1$\,nm and $a_2=10$\,nm and (d) $a_1=10$\,nm and $a_2=10.34$\,nm  compared to the SWCNT frequencies $\omega_1$ and $\omega_2$.}
\label{fig: Dispersion_modes}
\end{figure}

Figure \ref{fig: Dispersion_modes}(a) shows the dispersion curves $\omega_{\pm}(m,k)$ for the first modes for a DWCNT with $a_1=1$\, nm and $a_2=2$\,nm. The first mode $m=0$ may not satisfy the resonance condition for high velocities, while the modes with $|m|>0$ have always a solution $k_m^{\pm}$ for any velocities. Besides, it can be observed that the modes $\omega_{\pm}(m,k)$ converge for a sufficiently large value of the wavenumber $k$. Hence, if the resonance condition $k_m^{\pm}$ is sufficiently large, then all modes will have a similar value of $k_m^{\pm}$. It is worth noting that a large value of $k_m^{\pm}$ indicates that the associated wavelength of the wakefield ($\lambda_m^{\pm}=2\pi/k_m^{\pm}$) will be smaller. 

Moreover, the fundamental modes $\omega_{\pm}(0,k)$ are analysed in Fig. \ref{fig: Dispersion_modes}(b)-(d), since they are the only modes which contribute in the case of a driving particle travelling on axis (or if the wakefield is calculated on axis). Thus, Fig. \ref{fig: Dispersion_modes}(b) depicts the dependence of the fundamental mode $\omega_{\pm}(0,k)$ on the wavenumber $k$ for a DWCNT with radii $a_1=1$\,nm and $a_2=2$\,nm compared to the individual frequencies $\omega_1$ and $\omega_2$ associated with SWCNTs with radii $a_1$ and $a_2$, respectively. The dispersion relation $\omega_{+}$ is always higher than the individual frequencies $\omega_1$ and $\omega_2$, whereas $\omega_{-}$ is always smaller. Furthermore, $\omega_{+}$ exhibits a dependence on $k$ at small wavenumbers similar to the dispersion relation of a SWCNT, while $\omega_{-}$  exhibits a quasi-linear dispersion in the limit of small wavenumbers which can be expressed as $\omega_{-}(0,k\rightarrow 0)\approx v_\mathrm{p}k$ with $v_{\mathrm{p}}=\sqrt{4 \pi n_0\left[a_1 a_2 /\left(a_1+a_2\right)\right] \ln \left(a_2 / a_1\right)}$ \cite{MOWBRAY2005_MWCNT}. Resonance conditions are obtained in the intersection of $\omega_{\pm}$ and the velocity lines $kv$. If the velocity is too small the resonance conditions may not be satisfied. If the velocity increases both resonances $k_0^{\pm}$ will exist up to the velocity $v_\mathrm{p}$ when the resonance $k_0^{-}$ cannot be longer satisfied. If the velocity is sufficiently large the resonance condition $k_0^+$ may not be satisfied either.
Figure \ref{fig: Dispersion_modes}(c) shows the dispersion relation of the fundamental modes for $a_1=1$\,nm and $a_2=10$\,nm. As $a_2 \gg a_1$ the interaction between the fluids at each radius is very small and the dispersion modes $\omega_{\pm}(0,k)$ are similar to $\omega_1$ and $\omega_2$. However, as $\omega_{+}$ and $\omega_{-}$ are defined as the upper and lower curves, respectively, $\omega_1$ corresponds to $\omega_{-}$ up to the wavenumber $k^*$ that satisfies $\omega_1(k^*)=\omega_2(k^*)$ and for larger wavenumbers $\omega_1$ corresponds to $\omega_{+}$. Similarly, $\omega_2$ corresponds to $\omega_{+}$ up to the wavenumber $k^*$ while for larger wavenumbers corresponds to $\omega_{-}$. For completeness, in Fig. \ref{fig: Dispersion_modes}(d) the fundamental modes are depicted for a case with very close values for $a_1$ and $a_2$: $a_1=10$\,nm and $a_2=10.34$\,nm (0.34\, nm is the typical inter-wall distance in MWCNTs). Consequently, the frequency $\omega_{+}$ is similar to the individual frequency $\omega_1$ if calculated for a surface density $n_0=2n_g$, since this case can be approximated to a SWCNT with double density. It is also worth noting that if the DWCNT radii increase, the resonance condition can be satisfied for a wider range of high velocities, whereas if the surface density $n_0$ increases, then the dispersion curves increase and the resonance conditions are obtained for higher $k_m^{\pm}$. In general, a second resonance $k_m^{\pm}$ with a very high wavenumber may exist but the contribution from this resonance is normally totally negligible as it was concluded for a SWCNT in \cite{Martin-Luna2023_ExcitationWakefieldsSWCNT_NJP}.

\subsection{Electric wakefields}\label{Electric wakefields}

\begin{figure}[t!]
\includegraphics[width=\columnwidth]{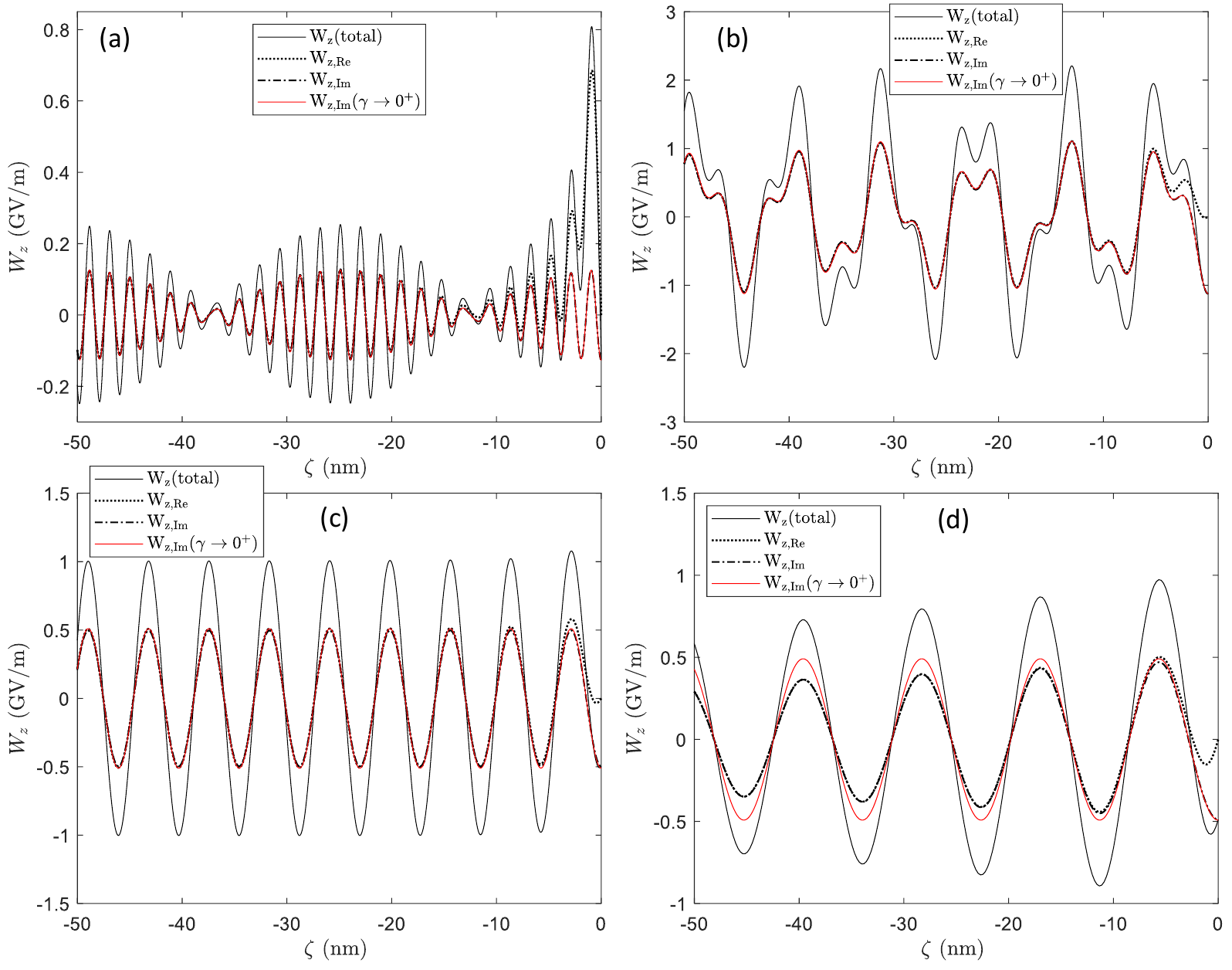}
\centering
\caption{Induced longitudinal wakefield contributions on axis ($r=0$) for a proton travelling on axis ($r_0=0$) at different velocity values: (a) $v=0.03c$, (b) $v=0.045c$, (c) $v=0.06c$ and (d) $v=0.09c$. The DWCNT radii are $a_1=1$\,nm and $a_2=2$\,nm and the friction parameter is $\gamma=10^{-4}\Omega$ in cases (a)-(c), and $\gamma=10^{-2}\Omega$ in (d), where $\Omega=\sqrt{4\pi n_0/a_1}$. The red curves show the approximation for small damping (Eq. (\ref{eq:Wz_im})). Note that the driving proton is at the comoving coordinate $\zeta=0$.}
\label{fig: Wakefield_terms}
\end{figure}


For the sake of simplicity, in the following calculations, it is assumed that the point-like charged particle is a proton, i.e. $Q=1$. Figure~\ref{fig: Wakefield_terms} shows the contributions of $W_{z,\text{Re}}$ and $W_{z,\text{Im}}$ to the induced longitudinal wakefield (cf. Eq. (\ref{eq:Wz})) for the velocities displayed at Fig. \ref{fig: Dispersion_modes}(b). On the one hand, Fig.~\ref{fig: Wakefield_terms}(a)-(b) shows a wakefield with two resonant frequencies because the velocities are smaller than $v_\mathrm{p}\approx 0.050c$. Figure~\ref{fig: Wakefield_terms}(a) shows a beat pattern which is typical when the resonant frequencies are not too different, whereas Fig.~\ref{fig: Wakefield_terms}(b) depicts a more complicated wakefield because the resonant frequencies are not close. On the other hand, Fig.~\ref{fig: Wakefield_terms}(c)-(d) shows a wakefield with a single resonant wavenumber: $k_0^+$, as it is deduced from Fig. \ref{fig: Dispersion_modes}(b). In Fig.~\ref{fig: Wakefield_terms}(d) the friction parameter has been increased to show that $\gamma$ produces an exponential decay of the induced wakefield with the distance behind the driving charged particle. It can be observed that the wavelength of the wakefield increases with the velocity $v$ as it was deduced from the dispersion relation. Furthermore, $W_{z,\text{Re}}$ and $W_{z,\text{Im}}$ are practically identical, except in the proximity of the driving particle. Consequently, when the damping parameter $\gamma$ is very small, the induced longitudinal wakefield can be approximated by $W_z\approx 2W_{z,\text{Im}}$ ($W_{z,\text{Im}}$ calculated using Eq. (\ref{eq:Wz_im})). The same approximation can be used for the transverse wakefield $W_r$.

\begin{figure}[!h]
\includegraphics[width=\columnwidth]{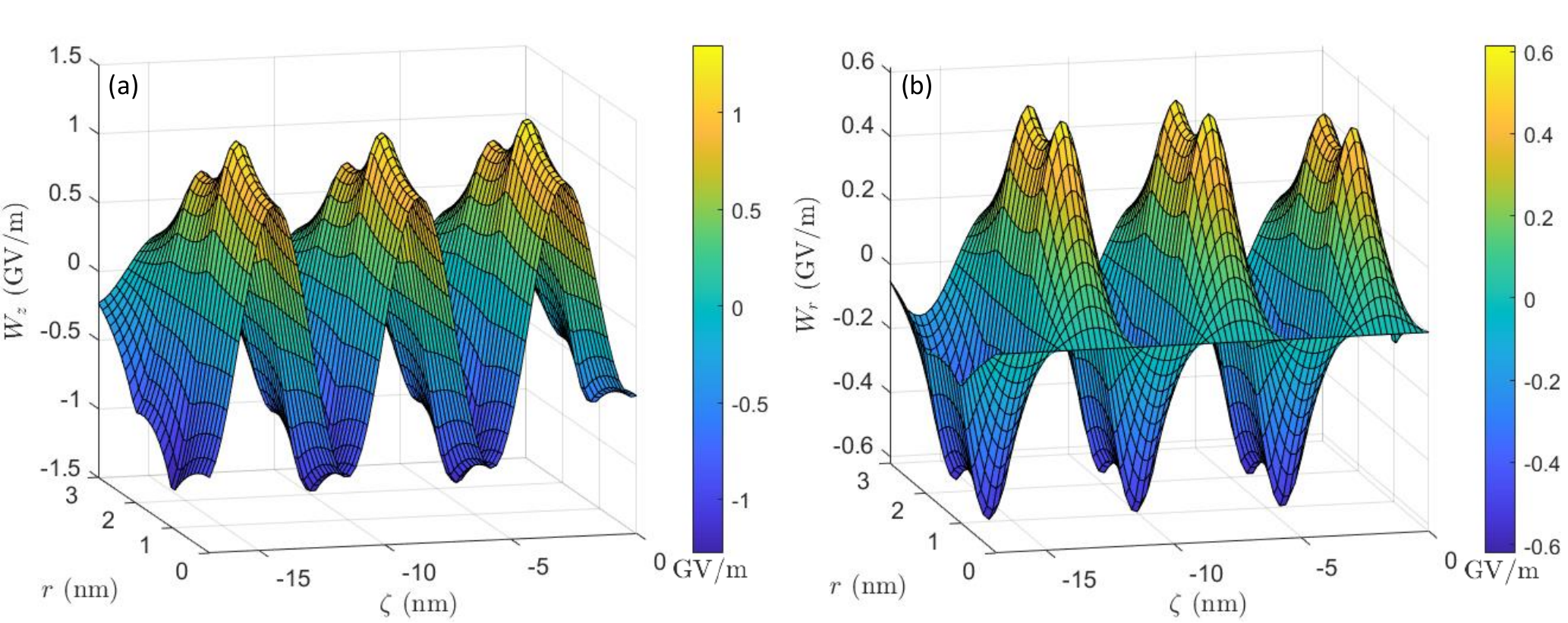}
\centering
\caption{Induced (a) longitudinal and (b) transverse wakefields in the $rz$-plane generated by a driving proton travelling on-axis inside a DWCNT considering the following parameters: $a_1=1$\,nm and $a_2=2$\,nm, $v=0.06c$ and $\gamma=10^{-4}\Omega$, where $\Omega=\sqrt{4\pi n_0/a_1}$.}
\label{fig:wakefields_inside_tube}
\end{figure}

Figure \ref{fig:wakefields_inside_tube} shows an example of the induced longitudinal and transverse wakefields in a DWCNT with $a_1=1$\,nm and $a_2=2$\,nm. It can be seen that the wakefield intensity increases near the cylinder walls. In  particular, if $r<a_1$, the wakefields increase with the radial distance $r$ because of the dependence on $I_m(|k|r)$ and $I_m^\prime(|k|r)$ (both increasing functions with the argument) of $W_z$ and $W_r$, respectively (cf. Eqs. (\ref{eq:Phi_ind_FB}), (\ref{eq:Wz}), (\ref{eq:Wr})). The same reasoning can be used to demonstrate that the wakefield intensity is higher if the particle travels off axis ($r_0\neq0$), although in this case higher order modes ($|m|>0$) should be computed. These effects will become more important for lower velocities, since the value of the resonance wavenumbers increase.

\begin{figure}[!h]
\includegraphics[width=\columnwidth]{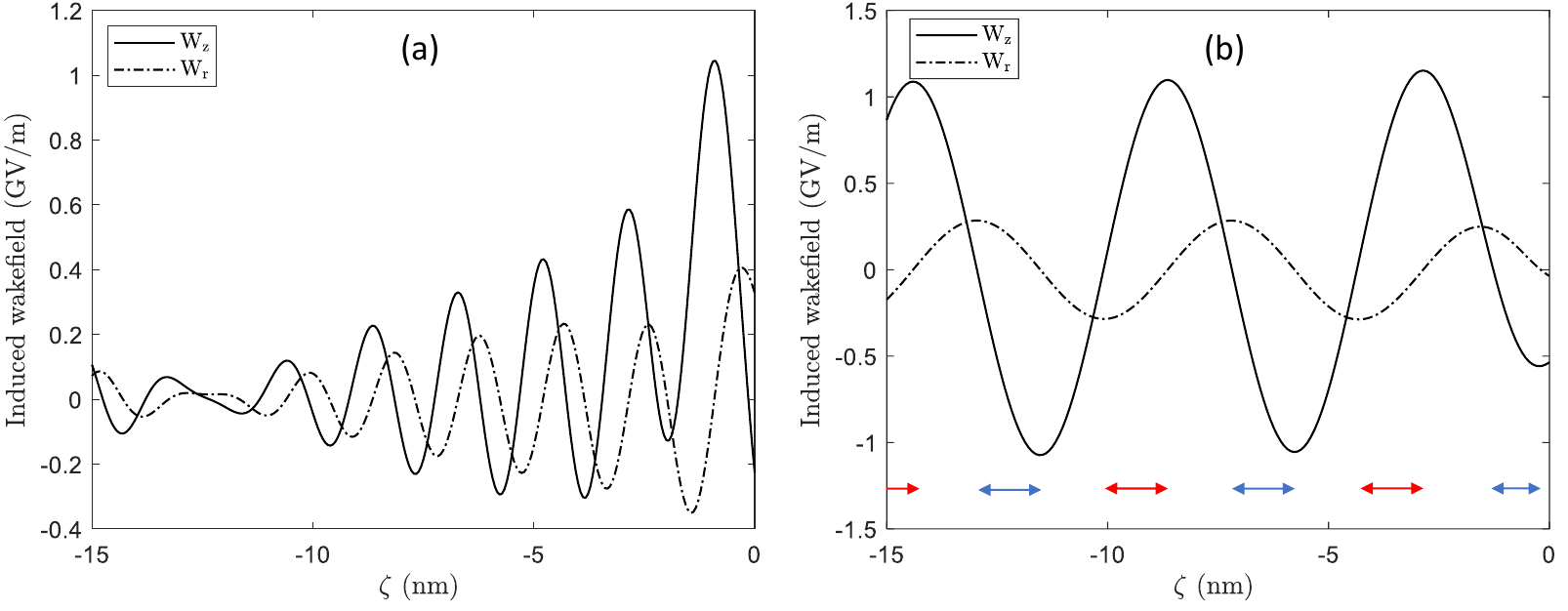}
\centering
\caption{Induced wakefields produced by a proton travelling on-axis with a velocity (a) $v=0.03c$ and (b) $v=0.06c$ at $r=a_1/2$. The DWCNT parameters are: $a_1=1$\,nm, $a_2=2$\,nm and $\gamma=10^{-4}\Omega$, where $\Omega=\sqrt{4\pi n_0/a_1}$. In (b) we have added red (blue) arrows indicating the regions where a positive (negative) witness charged particle would experience both acceleration and focusing simultaneously.}
\label{fig:wakefields_at_r05a}
\end{figure}

Furthermore, it is worth noting that there is a phase offset of $\pi/2$ between the induced longitudinal and transverse wakefields. Consequently, there are regions where the witness charged particles may simultaneously experience both acceleration and focusing (if they travel off-axis), similarly to what is observed for wakefields excited in homogeneous plasmas in the linear regime \cite{esarey2009physics_RevModPhys.81.1229} and in SWCNTs \cite{Martin-Luna2023_ExcitationWakefieldsSWCNT_NJP}. In DWCNTs, these regions will be periodical if there is a single resonant wavenumber (see Fig. \ref{fig:wakefields_at_r05a}(b)) or approximately periodical if the two resonant wavenumbers are excited (see Fig. \ref{fig:wakefields_at_r05a}(a)).

\subsection{Analysis of the excited modes}\label{Analysis of the excited modes}

\begin{figure}[!h]
\includegraphics[width=\columnwidth]{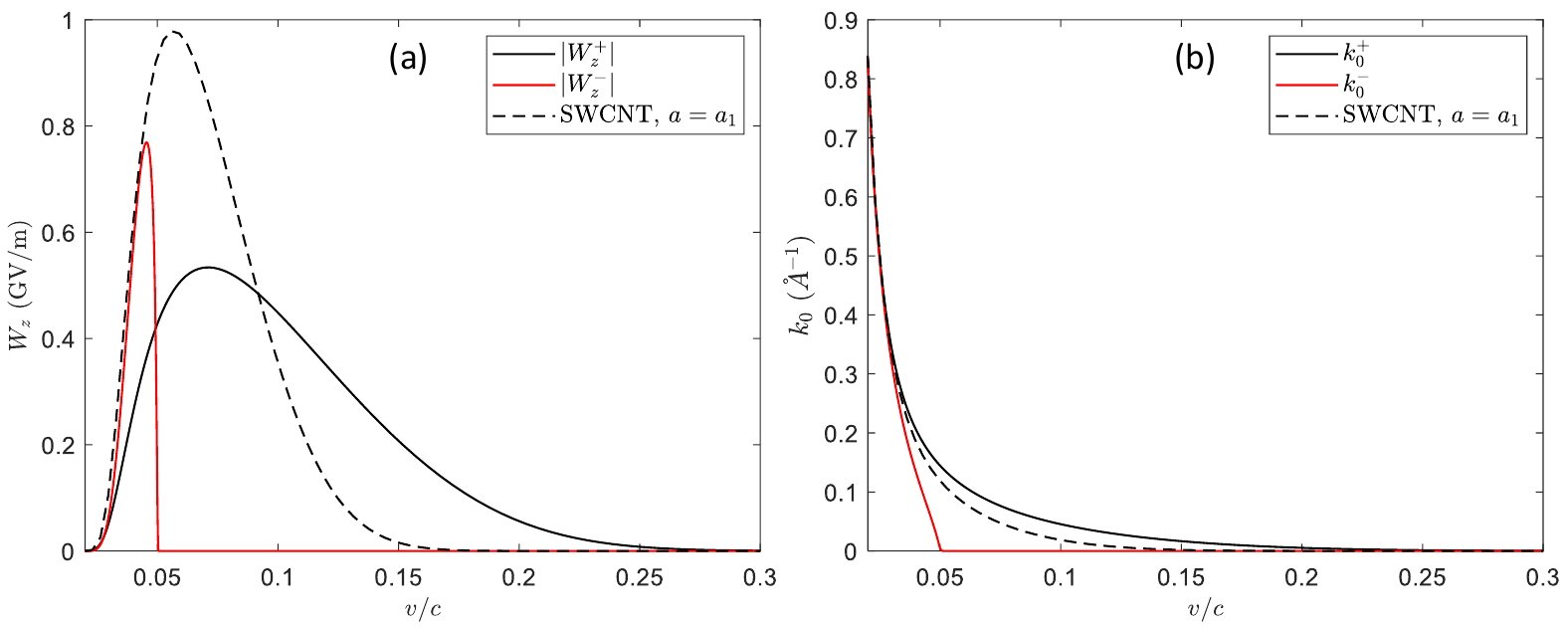}
\centering
\caption{(a) Amplitudes $W_z^{\pm}$ and (b) resonant wavenumbers $k_0^{\pm}$ as a function of the driving velocity. The parameters used are: $a_1=1$\,nm, $a_2=2$\,nm, $r=r_0=0$. The results are compared with the case of a SWCNT with radius $a=a_1$ (Eq. (15) in \cite{Martin-Luna2023_ExcitationWakefieldsSWCNT_NJP}).}
\label{fig:Wz2_pm_a1_2}
\end{figure}

As pointed out in the previous section, the plasmonic excitations can be approximated by $2W_{z,\text{Im}}$ when the damping parameter $\gamma$ converges to zero in the considered system. Therefore, we are going to analyze the contribution to $W_{z,\text{Im}}$ of the two possible resonant frequencies.  This section is focused on the plasmonic excitation created on axis by a proton travelling on axis, i.e. $r=r_0=0$. Hence, the only mode that contributes is $m=0$ and we will denote $W_z^{\pm} \equiv W^{\pm}_{z,0}$. Thus, Fig. \ref{fig:Wz2_pm_a1_2} depicts the absolute value of $W_z^{\pm}$ (because they are negative quantities for a proton) and the resonant wavenumbers $k_0^{\pm}$  as a function of the driving velocity compared to the results for a SWCNT with a radius $a=a_1$ (Eq. (15) in \cite{Martin-Luna2023_ExcitationWakefieldsSWCNT_NJP}). It can be seen that $W_z^-$ has a narrow peak at low velocities with a similar trend that the amplitude for a SWCNT until is cut at $v_p\approx 0.05c$, whereas $W_z^+$ has a broader peak because more velocities can produce a resonant excitation, as seen in Section \ref{Dispersion relation}. Moreover, the resonant wavenumbers decrease with the driving velocity, as it is deduced from the dispersion relation. It is worth noting that the stopping power is simply given by $S=|Q|(|W_z^+|+|W_z^-|)$ and, consequently, can easily be obtained from Figure \ref{fig:Wz2_pm_a1_2}(a), which shows a double-peak behaviour which agrees with the results obtained in \cite{MOWBRAY2005_MWCNT}. 

\begin{figure}[!h]
\includegraphics[width=\columnwidth]{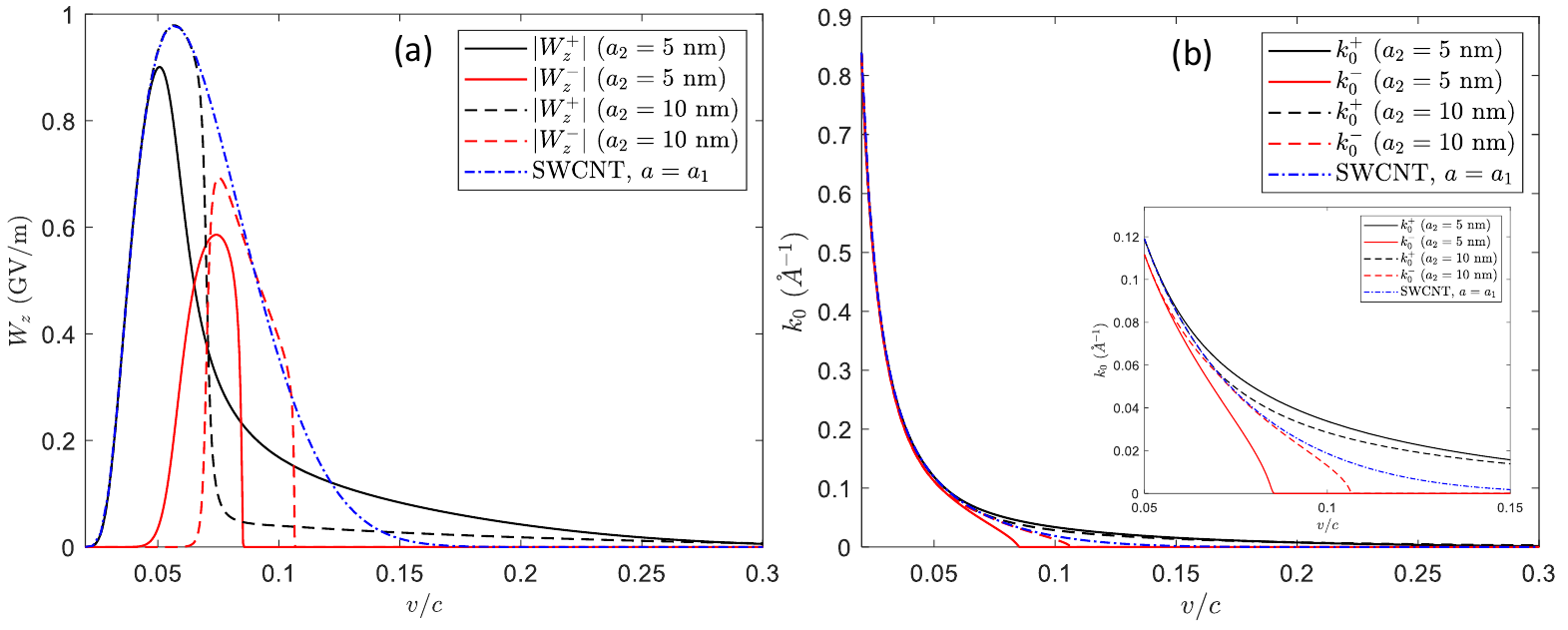}
\centering
\caption{(a) Amplitudes $W_z^{\pm}$ and (b) resonant wavenumbers $k_0^{\pm}$ as a function of the driving velocity for different $a_2$. The parameters used are: $a_1=1$\,nm and $r=r_0=0$. The results are compared with the case of a SWCNT with radius $a=a_1$ (Eq. (15) in \cite{Martin-Luna2023_ExcitationWakefieldsSWCNT_NJP}). In (b) is included a zoom of the region $0.05<v/c<0.15$.}
\label{fig:Wz2_pm_a2_5_10}
\end{figure}

If the external radius $a_2$ is increased the excited wakefields are similar to those obtained in a SWCNT with a radius $a=a_1$, as seen in Fig. \ref{fig:Wz2_pm_a2_5_10}. In particular, the peak produced in $W_z^-$ moves to higher velocities and widens because $v_p$ increases as $a_2$ grows. In an opposite way, the peak produced in $W_z^+$ moves to smaller velocities and narrows. If $a_2\gg a_1$ it can be observed that $W_z^+$ agrees with the curve for a SWCNT for low velocities, but at the velocity $v^*=\omega_1(k^*)/k^*$, $W_z^+$ falls to zero, and then $W_z^-$ increases tending to the SWCNT curve up to the velocity $v_p$ (which increases with $a_2$) when falls to zero as well. The transition at $k^*$ is more abrupt if $a_2$ increases. This behaviour is produced because, as it was seen in Fig. \ref{fig: Dispersion_modes}(c), when $a_2 \gg a_1$, $\omega_1$ corresponds to $\omega_{-}$ up to the wavenumber $k^*$ that satisfies $\omega_1(k^*)=\omega_2(k^*)$ and for larger wavenumbers $\omega_1$ corresponds to $\omega_{+}$. It can be seen that, in the range of velocities where $W_z^+$ ($W_z^-$) approaches to the wakefield of the SWCNT with radius $a=a_1$, the corresponding wavenumber $k_0^+$ ($k_0^-$) is similar to the SWCNT resonance $k_0$. This means that, in this case, the DWCNT can be approximated by a SWCNT with a radius $a=a_1$ and the really important dispersion relation is $\omega_1$. 

\begin{figure}[!h]
\includegraphics[width=\columnwidth]{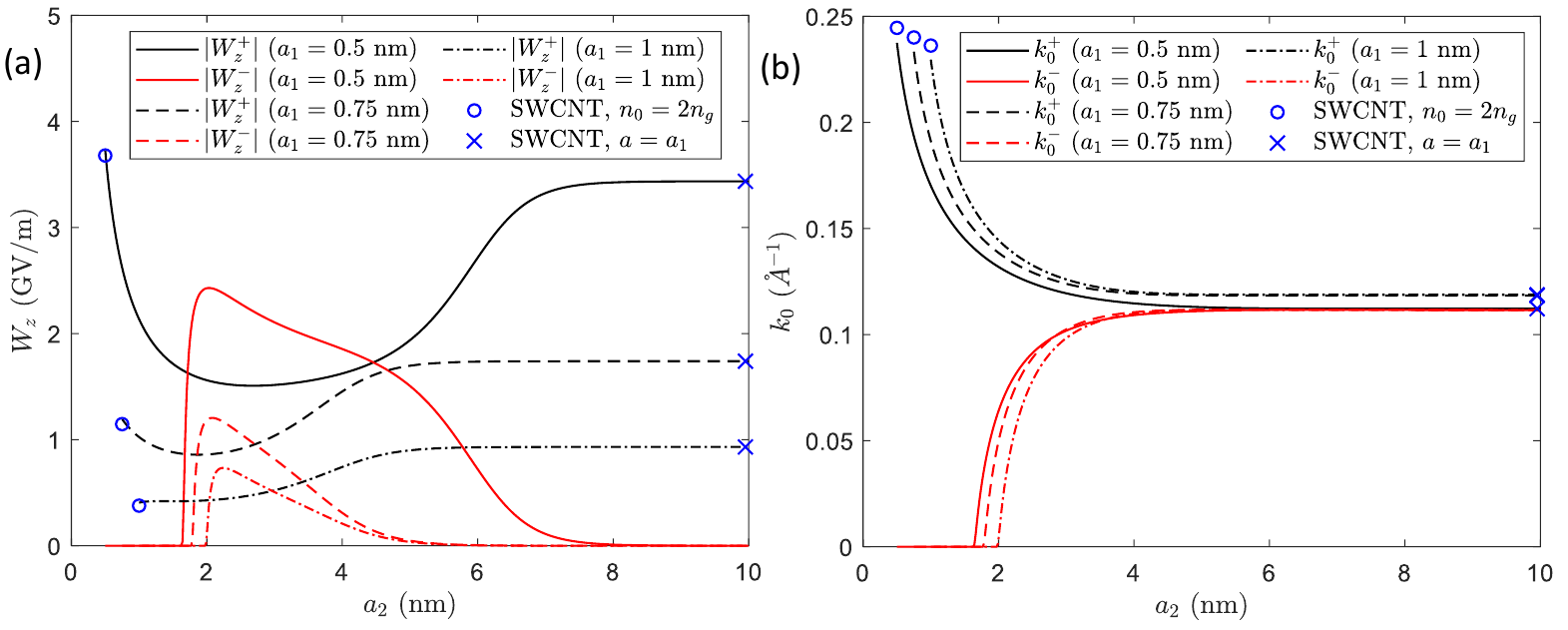}
\centering
\caption{(a) Amplitudes $W_z^{\pm}$ and and (b) resonant wavenumbers $k_0^{\pm}$ as a function of the external radius $a_2$ for different $a_1$. The parameters used are: $v=0.05c$ and $r=r_0=0$. The results are compared with the case of a SWCNT with $a=a_1$ and $n=2n_g$ (circles) and a SWCNT with $a=a_1$ and $n=2n_g$ (crosses) (Eq. (15) in \cite{Martin-Luna2023_ExcitationWakefieldsSWCNT_NJP}).}
\label{fig:Wz_scan_a2_DWCNT}
\end{figure}

Figure \ref{fig:Wz_scan_a2_DWCNT} shows the amplitudes $W_z^{\pm}$ and their corresponding wavenumbers as a function of the external radius $a_2$ for different internal radius $a_1$ and $v=0.05c$. If $a_2 \rightarrow a_1$ the $W_z^-$ mode is not excited and the $W_z^+$ mode is similar to the case of a SWCNT with $n_0=2n_g$ and radius $a=a_1$. These results agree with the analysis that was made of the dispersion relation of a DWCNT with two close cylinders (see Fig. \ref{fig: Dispersion_modes}(d)). The agreement with the case of a SWCNT with $n_0=2n_g$ is not exact because if $n_0$ changes, $\alpha$ is modified as well. If $a_2 \gg a_1$ the $W_z^-$ mode is not excited and the $W_z^+$ mode tends to a constant value in agreement with the excitation of a SWCNT with $n_0=2n_g$ and radius $a=a_1$, i.e., it is as if we removed the external cylinder with radius $a_2$. For intermediate values of $a_2$, the excitation of both $W^{\pm}_z$ modes may exist.

\begin{figure}[!h]
\includegraphics[width=\columnwidth]{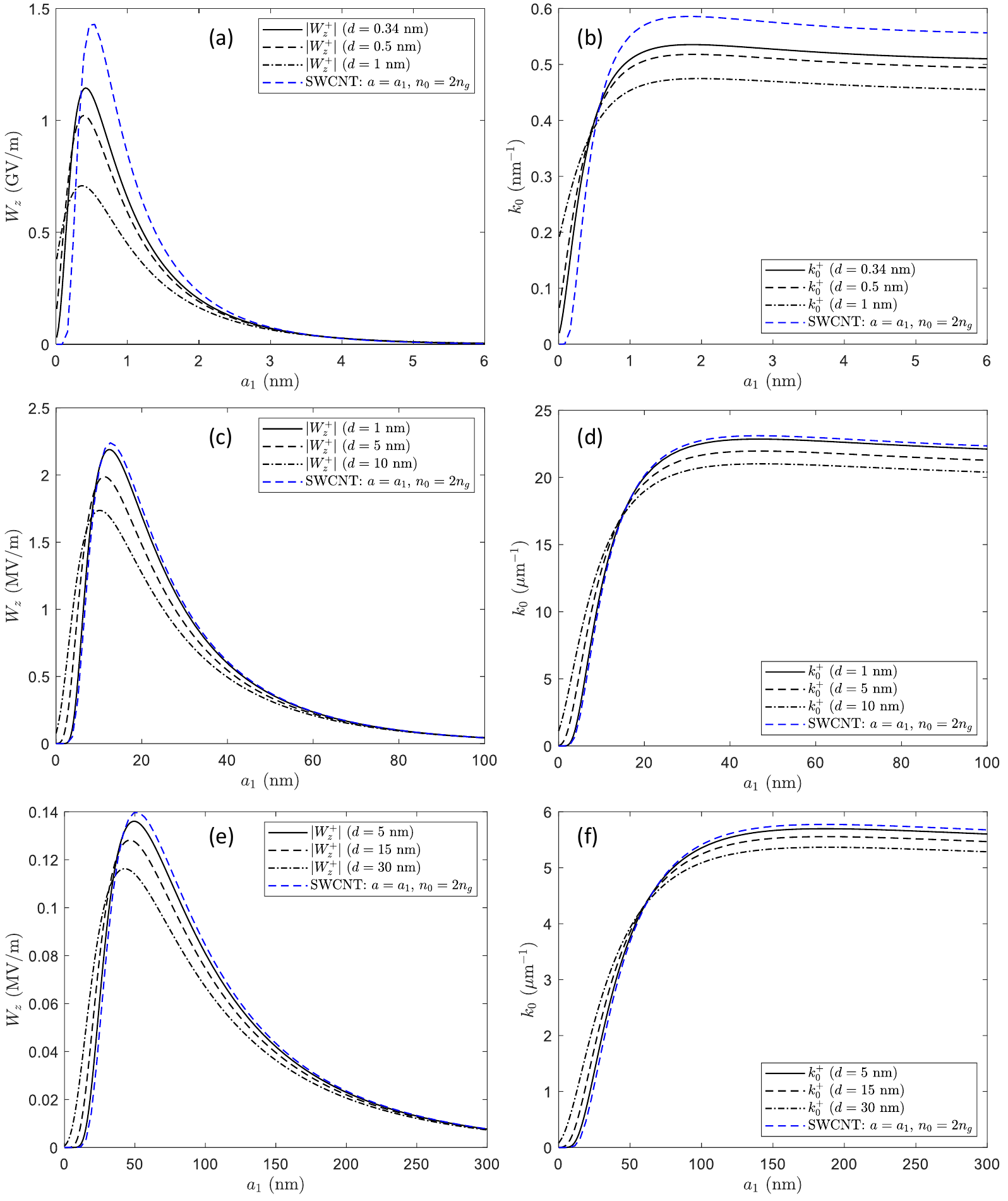}
\centering
\caption{Amplitude $W_z^+$ and resonant wavenumbers $k_0^+$ as a function of the internal radius $a_1$ for different inter-wall distances $d$ and different velocities: (a)-(b) $v=0.1c$, (c)-(d) $v=0.5c$ and (e)-(f) $v=c$. It is employed $r=r_0=0$. The results are compared with the case of a SWCNT with $a=a_1$ and $n=2n_g$ (Eq. (15) in \cite{Martin-Luna2023_ExcitationWakefieldsSWCNT_NJP}).}
\label{fig:interwall_distance_DWCNT}
\end{figure}

Finally, we are going to analyze the excitation of the wakefields in a DWCNT with a constant inter-wall distance $d$ (i.e. $a_2=a_1+d$). Thus, Fig. \ref{fig:interwall_distance_DWCNT} shows the amplitude $W_z^+$ ($W_z^-$ is negligible) as a function of the internal radius $a_1$ for different inter-wall distances $d$ and driving velocities $v$. It can be seen that for higher velocities the excitation of both $W^{\pm}_z$ modes may exist at higher internal radius, as it happens in SWCNTs \cite{Martin-Luna2023_ExcitationWakefieldsSWCNT_NJP}. On the other hand, as the inter-wall distance decreases, the curves converge to the case of a SWCNT with $n_0=2n_g$. This approximation is better if $d/a_1 \ll 1$ and, consequently, for typical inter-wall distances (e.g. $d=0.34$\,nm) it works better for larger $a_1$ which are the radius needed for ultra-relativistic driving velocities. In consequence, DWCNTs with large $a_1$ and small $d$ can be optimised to obtain the highest wakefields employing the expressions for SWCNTs described in \cite{Martin-Luna2023_ExcitationWakefieldsSWCNT_NJP}, but using $n_0=2n_g$. Such as in the case of SWCNTs with higher surface density $n_0$, in which larger wakefields can be achieved \cite{Martin-Luna2023_ExcitationWakefieldsSWCNT_NJP}, DWCNTs with small inter-wall distances can be used to obtain higher wakefields.

\section{Conclusions}\label{Conclusions}

The linearized hydrodynamic model has been employed to describe the plasmonic excitations generated by a point-like charge moving parallel to the axis in a DWCNT. We have derived general expressions for the longitudinal and transverse wakefields. Their dependencies on the radii of the DWCNT and the velocity of the driving charged particle have been numerically studied and related to the dispersion relation. If the friction parameter is negligible, the plasmonic excitations can be approximated by twice the Equations (\ref{eq:Wz_im})-(\ref{eq:Wr_im}). In these equations two different resonant wavenumbers appears because the dispersion relation of a DWCNT splits in two branches compared to the case of a SWCNT. Thus, a particle travelling on-axis may excite two different modes which have been analysed in detail using the Equations (\ref{eq:Wz_mas})-(\ref{eq:Wz_menos}) and can clearly explain the stopping power obtained in previous works in DWCNTs \cite{MOWBRAY2005_MWCNT, YOU2009_KineticModelDWCNT_NIMB}. It has been shown that if the external radius is far from the internal radius, the obtained wakefields are similar to those of a SWCNT with a radius equal to the internal radius of the DWCNT. Furthermore, if the inter-wall distance is much smaller than the internal radius the excited wakefields are similar to the case of a SWCNT with double the surface density. Therefore, DWCNTs can be an option to obtain a system that behaves as SWCNTs with higher surface electron densities which, consequently, can provide greater longitudinal wakefields that can be optimised using simpler equations \cite{Martin-Luna2023_ExcitationWakefieldsSWCNT_NJP}. In particular, it can be interesting for ultra-relativistic driving particles since they produce the highest wakefields in conventional SWCNTs with radius in the order of $\sim100$ nm and the periodic regions where a witness beam can simultaneously experience acceleration and focusing are wider so obtaining these beams will be easier. Thus, the excitation of these plasmonic modes with ultra-high gradients can be employed for particle acceleration or to produce coherent radiation since electrons experience betatron motion along the tube \cite{Shin2017OpticallyControlledCoherentX-rayRadiationsFromPhotoExcitedNanotubes}.

\section*{CRediT authorship contribution statement}
\textbf{Pablo Mart\'in-Luna:} Conceptualization, Methodology, Software, Validation,  Formal Analysis, Investigation, Writing - original draft, Writing - review \& editing. \textbf{Alexandre Bonatto:} Supervision, Writing - review \& editing. \textbf{Cristian Bontoiu} Supervision, Writing - review \& editing. \textbf{Guoxing Xia:} Supervision, Writing - review \& editing. \textbf{Javier Resta-L\'opez:} Conceptualization, Methodology, Validation, Formal analysis, Investigation, Resources, Writing - original draft, Writing - review \& editing, Supervision, Project Administration,  Funding acquisition.

\section*{Declaration of competing interest}
The authors declare that they have no known competing financial interests or personal relationships that could have appeared to influence the work reported in this paper.

\section*{Data availability}
Data will be made available on request.

\section*{Acknowledgments}
This work has been supported by Ministerio de Universidades (Gobierno de Espa\~{n}a) under grant agreement FPU20/04958, and the Generalitat Valenciana under grant agreement CIDEGENT/2019/058. 

\section*{References}
\bibliography{References}
\providecommand{\noopsort}[1]{}\providecommand{\singleletter}[1]{#1}%
\providecommand{\newblock}{}

\end{document}